\documentstyle[11pt,moriond_pre,epsfig]{article}

\begin{document}
\begin{flushright}
FERMILAB-CONF-98/242-T \\
hep-ph/9808279
\end{flushright}
\vspace*{3cm} \title{THREE ELECTROWEAK RESULTS FROM LATTICE QCD
\footnote{Talk presented at the XXXIIIrd Rencontres de Moriond on
``Electroweak Interactions and Unified Theories,''
14--21 March 1998, Les Arcs, France.}
}

\author{Andreas S.~KRONFELD}

\address{Theoretical Physics Department, Fermi National Accelerator
Laboratory, Batavia, Ill., USA}

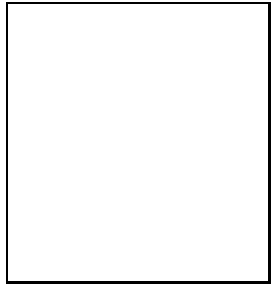
\maketitle
\vfill
\abstracts{Quantum chromodynamics is needed to understand quarks and,
hence, to determine the quarks' Yukawa couplings from experimental
measurements.
As a short illustration, the results of three lattice calculations
are given.}
\abstracts{La Chromodynamique Quantique est n\'ec\'essaire pour
comprendre le comportement des quarks et, en cons\'equence, pour
d\'eterminer les constantes de couplage de Yukawa des quarks \`a partir
de mesures exp\'erimentales.
En guise d'illustration extr\^emement courtes, les r\'esultats de trois
calculs sur r\'eseau sont pr\'esentes.}

\vfill
\newpage

\section{Introduction}
The crudest model of electroweak symmetry breaking couples quarks to
a Higgs doublet~$\Phi$ as follows:
\begin{equation}
{\cal L}_Y=
\left(\bar{d} \;\; \bar{s} \;\; \bar{b} \right)_R
\left(\begin{array}{c} {\rm some} \\
	{\rm Yukawa} \\
	{\rm \!couplings\!} \end{array}\right)
\left(\!\begin{array}{c} \Phi^\dagger Q^1_L \\
	\Phi^\dagger Q^2_L \\
	\Phi^\dagger Q^3_L \end{array}\!\right) +
\left(\bar{u} \;\; \bar{c} \;\; \bar{t} \right)_R
\left(\begin{array}{c} {\rm more} \\
	{\rm Yukawa} \\
	{\rm \!couplings\!} \end{array}\right)
\left(\!\begin{array}{c} \bar{\Phi} Q^1_L \\
	\bar{\Phi} Q^2_L \\
	\bar{\Phi} Q^3_L \end{array}\!\right) +~{\rm h.c.}
    \label{Yukawa}
\end{equation}
In the phase with spontaneously broken symmetry, $\Phi$ obtains a vacuum
expectation value $\langle\Phi^\dagger\rangle=(0\;\;v)$.
Then the eigenvalues of the Yukawa-coupling matrices, times~$v$, give
the six quarks a mass.
The observable part of the eigenvectors boils down to the
Cabibbo-Kobayashi-Maskawa (CKM) mixing matrix, a four-parameter object.
Thus, the Yukawa-coupling matrices provide ten of the nineteen free
parameters of the Standard Model.

The structure in~(\ref{Yukawa}) is ugly and begs an explanation.
The large number of free parameters is only a superficial scar.
At a deeper level, quantum field theories with simple scalar fields,
such as~$\Phi$ in~(\ref{Yukawa}), are unlikely to be consistent at
higher and higher energies.
It is more likely that~$\Phi$ is just a placeholder for Nature's way
of breaking electroweak symmetry.
Eventually, experiments will reveal how Nature does it.

In the absence of experimental data, theorists like to extend
the Standard Model with elegance as a guide.
Usually that means a few more fields but much more symmetry.
Even experimenters cannot resist these ideas, so they devote much of
their resources to the search for partners postulated by supersymmetry,
vector bosons associated with unifying gauge groups, and scalar
bosons---besides~$\Phi$---predicted by these and other extensions.

A complement to direct searches for, say, squarks and gluons, is to pin
down the Standard Model more precisely.
A convenient way to summarize experiments relevant to~(\ref{Yukawa}) is
through the six quark masses and the four parameters of the CKM matrix.
Then you can trace back through the Yukawa couplings to the Lagrangian
of your favorite supersymmetric, grand-unified, theory of everything.
Unfortunately, when the data are presented this way we do not
seem to know much.
The underlying reason is mostly theoretical:
the quarks' color is confined into hadrons, so to deduce the
properties of quarks, one must face the theory of the strong
interactions: quantum chromodynamics (QCD).
And not just perturbative QCD, but QCD with all its nonperturbative
challenges.

This talk presents three recent results:
the masses of the light quarks (including strange),\cite{Gou97}
the mass of the charmed quark,\cite{Kro98} and
the decay constants of the $B$ and $D$ mesons.\cite{Rya98}
Although QCD, formulated on a lattice, is used as a tool,\cite{Kro93}
the motivation is mostly electroweak for the masses and entirely
electroweak for the decay constants.

The calculations presented here were performed on ACPMAPS, a parallel
supercomputer designed at Fermilab.
In each case we have carried out the needed computations with an action
designed to remove the leading cutoff effect.
We also employ three different lattice spacings to control cutoff
effects further.
This means that the numbers reported here are meant to reflect
{\em continuum\/} QCD, and the error bars reflect our confidence in
exptrapolating lattice artifacts away.

The main shortcoming of the results is that we have used quenched QCD,
rather than full QCD.
(Quenched QCD omits the back-reaction of vacuum quark loops on the
gluons.)
Our quoted uncertainties attempt also to account for the error made
here, but it must be acknowledged that this part of the error analysis
is somewhat subjective.

Owing to lack of space, this talk is not a review.
Thus, remarks on and references to other work is selective;
readers interested in more details should look elsewhere.

\section{Results}

\subsection{Strange and Light Quarks' Masses}
In QCD each quark mass is a free parameter that can be chosen to fit
(one) experiment.
They are then fixed while checking other predictions of QCD.
When the light and strange quark masses are chosen to fit $m_\pi$ and
$m_K$, the other meson and baryon masses of quenched lattice QCD are in
fair, but not perfect, agreement.
The discrepancy remains in the continuum limit and is, thus, most likely
an artifact of the quenched approximation.

The quark mass that comes out of such an analysis is not 
an inertial or gravitational mass, but, instead, a running mass in a
scheme that depends on details of the calculation.
It is possible to convert the lattice mass to other schemes,
usually with the help of perturbation theory.
It has become customary to quote quark masses in the $\overline{\rm MS}$
scheme at the renormalization point 2~GeV.
We find, for the average of the light quarks and for the strange
quark,\cite{Gou97}
\begin{equation}
\bar{m}_l(2~{\rm GeV}) = 3.6(6)~{\rm MeV}, \quad
\bar{m}_s(2~{\rm GeV}) = 95(16)~{\rm MeV},
\label{light}
\end{equation}
in the quenched approximation.
The uncertainties are dominated by the extrapolation to the continuum
limit, and then by the (one-loop) perturbative conversion
to~$\overline{\rm MS}$.
These results are lower than the (pre-1997) conventional wisdom, but
they are consistent with all other lattice results that extrapolate to
the continuum limit.\cite{Gup97,Bha97}

A criticism of this result, voiced also here at Moriond, is that the
conversion to $\overline{\rm MS}$ is done at one loop.
It would be better---of course---to convert nonperturbatively the
lattice mass to an intermediate scheme.
In one implementation of this idea\cite{Mar95} the resulting masses
at {\em nonzero\/} lattice spacing are rather larger.\cite{Gim98}
It seems, however, that the resulting lattice artifacts are large, and
an extrapolation to the continuum limit agrees very well
with~(\ref{light}).\cite{Bha98}
In the near future one can expect the {\sl Alpha\/}
Collaboration\cite{Cap98} to combine their nonperturbative mass
renormalization with hadron spectrum calculations to give a bullet-proof
determination of the quark mass, albeit in the quenched approximation.

What can one expect when the calculations are repeated in full QCD?
We have argued that the results should be 20--40\% lower
still:\cite{Gou97}
\begin{equation}
2.1~{\rm MeV} < \bar{m}_l(2~{\rm GeV}) < 3.5~{\rm MeV}, \quad
 54~{\rm MeV} < \bar{m}_s(2~{\rm GeV}) < 92~{\rm MeV}.
\label{nf}
\end{equation}
This estimate is based on early calculations with two flavors' worth of
vacuum quark loops.
In the meantime, more extensive calculations in full QCD have been
started, and preliminary results appear to substantiate
our estimate.\cite{Bur98}

\subsection{Charmed Quark's Mass}
We preliminarily find\cite{Kro98}
\begin{equation}
\bar{m}_{\rm ch}(m_{\rm ch}) = 1.33 \pm 0.08~{\rm GeV}.
\label{charm}
\end{equation}
Here the uncertainty stems mostly from perturbation theory, whereas
in~(\ref{nf}) uncertainties from quenching and
the extrapolation to the continuum limit dominate.
For a heavy quark, these two are less important.
The error introduced by quenching arises because the running mass
runs too quickly in the quenched approximation.\cite{Mac94}
The running stops at the threshold, so, renormalizing at threshold, a
heavy quark's mass should not depend (much) on quenching.\cite{Dav94}
The extrapolation is under better control here, because there are two
ways to define the quark mass with opposite lattice artifacts but the
same continuum limit.\cite{Kro98}
As before the perturbative calculation is available at the one loop
level only, so it will be difficult to reduce the uncertainty much
further.

\subsection{$B$ and $D$ Meson Decay Constant}
The decay constant of heavy-light meson describes the strong
interactions' effect on the decay vertex of the meson into leptons.
Thus, it is the simplest interesting matrix element of the electroweak
Hamiltonian.
Our results for the decay constants are\cite{Rya98}
\begin{eqnarray}
f_{D_d} = 194^{+14}_{-10}\pm 10~{\rm MeV}, & \quad &
\label{fD}
f_{D_s} = 213^{+14}_{-11}\pm 11~{\rm MeV}, \\
f_{B_d} = 164^{+14}_{-11}\pm\;8~{\rm MeV}, & \quad &
\label{fB}
f_{B_s} = 185^{+13}_{-08}\pm\;9~{\rm MeV}.
\end{eqnarray}
The first uncertainty is statistical and the second is from
perturbation theory.
The effects of the quenched approximation may raise these quenched
results by as much as 10\%.
These results agree with other modern
calculations.\cite{Aok98,Ali98,Ber98}

The leptonic decays $D_d\to\mu\nu$ and $D_s\to\mu\nu$ are frequent
enough to measure in high-statistics charm experiments.
Combining these measurements with reliable theoretical calculations
of~$f_D$ and~$f_{D_s}$ will yield a determination of the CKM matrix
elements~$|V_{cd}|$ and~$|V_{cs}|$.
Because a helicity flip suppresses the pure leptonic decay, semileptonic
decays (combined with their own QCD matrix elements) may prove better,
but, especially for those bent on extending it, more 
checks of the Standard Model are always good.

Leptonic decays of $B$ mesons have not yet been detected.
But the decay constants of neutral $B$s are used also to parametrize
mixing.
There the hadronic matrix element is $(8/3)m_B^2f_B^2B_B$, and $B_B$ is
within 10\% or so of unity.
Mixing between $B^0_d$ and $\bar{B}^0_d$ ($B^0_s$ and $\bar{B}^0_s$)
is also sensitive to $|V_{td}|$ ($|V_{ts}|$).
Thus, measurements of the mixing together with theoretical calculations
of the decay constants $f_B$ and bag parameters $B_B$ will pin down
these poorly known elements of the CKM matrix.

\end{document}